\documentclass[aps,pra,twocolumn,groupedaddress,showpacs]{revtex4}
\usepackage{bm}

\begin{document}

\def\ba{\mathbf{a}}
\def\d{\mathbf{d}}
\def\P{\mathbf{P}}
\def\bK{\mathbf{K}}
\def\bk{\mathbf{k}}
\def\bkn{\mathbf{k}_{0}}
\def\bx{\mathbf{x}}
\def\bfn{\mathbf{f}}
\def\bg{\mathbf{g}}
\def\bj{\mathbf{j}}
\def\bR{\mathbf{R}}
\def\br{\mathbf{r}}
\def\bu{\mathbf{u}}
\def\bq{\mathbf{q}}
\def\bw{\mathbf{w}}
\def\bp{\mathbf{p}}
\def\bG{\mathbf{G}}
\def\bQ{\mathbf{Q}}
\def\bs{\mathbf{s}}
\def\E{\mathbf{E}}
\def\bv{\mathbf{v}}
\def\b0{\mathbf{0}}
\def\la{\langle}
\def\ra{\rangle}
\def\beq{\begin{equation}}
\def\eeq{\end{equation}}
\def\bea{\begin{eqnarray}}
\def\eea{\end{eqnarray}}
\def\bdm{\begin{displaymath}}
\def\edm{\end{displaymath}}
\def\bnab{\bm{\nabla}}

\title{Two-fluid hydrodynamic modes in a trapped
superfluid gas}

\author{E. Taylor and A. Griffin}
\affiliation{Department of Physics, University of Toronto, Toronto, Ontario,
Canada, M5S 1A7}

\date{\today}

\begin{abstract}
In the collisional region at finite temperatures, the collective modes of
superfluids are described by the Landau two-fluid hydrodynamic equations.  This
region can now be probed over the entire BCS-BEC crossover in trapped Fermi
superfluids with a Feshbach resonance, including the unitarity region.  Building
on the approach initiated by Zaremba, Nikuni, and Griffin in 1999 for
trapped atomic Bose gases, we present a new variational formulation of
two-fluid hydrodynamic collective modes based on the work of Zilsel in 1950
developed for superfluid helium. 
Assuming a
simple variational ansatz for the superfluid
and normal fluid velocities, the frequencies of the hydrodynamic modes are given
by solutions of coupled algebraic equations, with constants only involving
spatial integrals over various equilibrium thermodynamic derivatives.  This
variational approach is both simpler and more physical than a direct
attempt to solve the Landau two-fluid differential equations.  Our two-fluid
results are shown to reduce to those of Pitaevskii and Stringari for a pure
superfluid at $T=0$.   
\end{abstract}

\pacs{03.75.Kk,~03.75.Ss,~67.40.-w}

\maketitle

\section{Introduction}

In 1938, Tisza~\cite{Tisza} first suggested that the classic features of
superfluid helium were a manifestation of a two-fluid
hydrodynamics, originating from a Bose-Einstein condensate.  The correct
two-fluid equations in the non-dissipative limit were formulated and
solved by Landau a few years later~\cite{Landau41,Khalatnikov}. Two-fluid hydrodynamics describes the coupled
dynamics of
the superfluid and normal fluid components when collisions produce a state of local thermodynamic
equilibrium.  This collision-induced state is described by the usual
thermodynamic variables, which are now dependent on position and time.  The
two-fluid equations are essentially the same as the fluid dynamics of a normal
fluid, but
with additional equations describing the new superfluid component.  In
simple terms they describe the collective oscillations of frequency $\omega$
such that $\omega \tau_{R} \ll 1$, where $\tau_R$ describes the time it takes
for a nonequilibrium state to reach local equilibrium (see, for example,
Ref.~\cite{Griffin01}). For brief accounts of
two-fluid hydrodynamics in the context of uniform quantum gases, see
Refs.~\cite{Pitaevskiibook, Pethickbook}.

The two-fluid regime was difficult to reach in the first wave of
experiments on trapped Bose-condensed gases after 1995.  The typical
collisional cross-section and the achievable densities were not sufficient to
reach the region described by Landau's two-fluid hydrodynamics.
An exception was the pioneering experiment by Ketterle and
coworkers~\cite{Ketterle98}, which later theoretical work showed was well within
the two-fluid region~\cite{Griffin01}.  However, there is new interest in the
study of the two-fluid collisional dynamics in trapped gases.  One reason is
that with atom chips, one can now produce tightly localized high density atomic
Bose condensates.  In addition, in a trapped two-component Fermi gas, one can
use a
Feshbach resonance to adjust the magnitude and sign of the s-wave scattering
length $a$ between Fermi atoms prepared in two different hyperfine states. 
This allows one to study the collective modes across the BCS-BEC crossover
region in great detail~(see, for example, Refs.~\cite{Thomas04,Grimm04}), with
extremely
large values of $|a|$.  One also expects that even at the unitarity limit
($|a|\rightarrow \infty$) of trapped Fermi superfluids, one is dealing with a
region where the Landau two-fluid equations are now the appropriate description
of the
dynamics at finite temperatures.  Several recent papers~\cite{Ho04, 
Heiselberg05} have considered first and second sound in a uniform
two-component Fermi gas in the
unitarity region. 

So far, discussions of collective modes in a trapped Fermi gas with strong
interactions have been limited to either $T=0$, where one is dealing
with a
pure
superfluid~\cite{Stringari04EPL, Heiselberg04, Zubarev04, Tosi04, Bulgac05,
Manini05,
Stringari05}, or to $T>T_{c}$, where one is dealing
with the hydrodynamics of a normal Fermi liquid~\cite{Bruun99,Pedri03,Smith05}.
The
much
richer collective mode spectrum at intermediate temperatures described by
Landau two-fluid
hydrodynamics awaits exploration.  In particular, the BEC side of the BCS-BEC
crossover is well described as a strongly interacting Bose gas of very stable
molecules~\cite{PetrovMol}.  This should be an ideal system to study the
analogue of first and second sound in a uniform Bose superfluid
predicted by the Landau two-fluid
equations.  

The Landau two-fluid equations describe \textit {all}
superfluids with a two-component order parameter in
terms of coupled differential equations for the local hydrodynamic variables and
the velocity fields of the two fluids.  The coefficients in these equations are
given in terms of equilibrium thermodynamic functions.  Thus, the only
differences between the hydrodynamics of various superfluids are the values of
these thermodynamic
functions.  These depend on the thermal excitations
which are
appropriate for different systems (superfluid helium, BCS superconductors,
trapped Bose gases, etc.).  The two-fluid equations for a BCS
superfluid have been derived by Galasiewicz~\cite{Galasiewicz}.

The usual derivation~\cite{Khalatnikov} of the Landau two-fluid hydrodynamic
equations is based on
conservation laws and macroscopic considerations, with key equations describing
the irrotational superfluid velocity field $\bv_s(\br, t)$ being postulated. 
There
is a
considerable literature on this approach, which makes little contact
with any microscopic theory of the origin of the superfluid
component. 
Since the 1960s, however, it has generally been accepted that this originates
from
Bose-Einstein condensation (for a review, see Ref.~\cite{Griffinbook}).  

In the context of \textit{trapped}
atomic gas superfluids, an important development was the work of Zaremba,
Nikuni,
and
Griffin (ZNG) in 1999~\cite{ZNGJLTP}, who discussed a simple microscopic model
of a condensate (described by a generalized Gross-Pitaevskii equation) coupled to a
thermal gas of atoms moving in a self-consistent Hartree-Fock (HF) field
(described by a kinetic Boltzmann equation generalized to include the Bose
statistical factors). 
The ZNG coupled equations for the condensate (superfluid)
and
non-condensate (normal fluid) have been used to discuss the collisionless
(see, for example, Ref.~\cite{accidentalZJ} and references therein) as well as
the
collision-dominated regions. In the latter region, ZNG showed that in the limit
that collisions
establish \textit{complete} local equilibrium ($\omega \tau_{R}\rightarrow 0$),
their equations can be reduced precisely to the Landau two-fluid
equations, with all
thermodynamic functions being given explicitly for a trapped gas of thermal atoms with a HF spectrum.  This formalism is immediately applicable to a molecular Bose condensate at finite
$T$, on the BEC side of a Feshbach resonance~\cite{Grimm04}. 

For a uniform superfluid, the
linearized Landau equations (without dissipation due to transport processes) are
solved in the standard texts~\cite{Khalatnikov, LL, Putterman,
Pethickbook, Pitaevskiibook}
and predict two
sound modes.  These are the in-phase (first sound) and out-of-phase (second
sound) oscillations of the two fluids, familiar in studies of bulk superfluid
helium.  The mode frequencies have been calculated for a uniform 
weakly-interacting
Bose-condensed gas (see Ref.~\cite{ZG97} and Appendix C in Ref.~\cite{NG01}). 
For a trapped superfluid, in
contrast, it is
very difficult to directly solve the Landau equations for the normal modes of
the gas~\cite{ZNGJLTP, Ho98}.  The present paper deals with a new approach for
calculating the frequencies of
the normal modes
of a trapped superfluid.

We develop a 
variational
approach of the kind first used by ZNG (for a slightly modified version of
the two-fluid equations reviewed in Appendix B).  Our discussion is based on an
action given in
terms of the velocity and density fields of the superfluid and normal
fluid, first proposed by Zilsel in 1950~\cite{Zilsel50}.  The stationary point
of the action corresponds to the solutions of the Landau
two-fluid differential
equations.  Considering fluctuations of this action about an equilibrium state,
we introduce simple variational ansatzes for the velocity fields.  The end
result is that the two-fluid hydrodynamic modes can be described in terms of
two coupled harmonic oscillators.  The effective masses and spring constants
of the two oscillators are given explicitly in terms 
of spatial integrals over various static equilibrium thermodynamic functions
and their derivatives.  The latter can be computed in a straightforward manner,
once we have a specific model for the thermal excitations of the trapped
superfluid. 

We illustrate our formalism in Section VI by using it to derive first and
second sound modes in a uniform superfluid as well as the frequencies of
the
dipole and breathing modes of a trapped superfluid.  

There is an extensive older literature~\cite{Zilsel50, Yourgau, Lhuillier75,
Geurst76, Jackson78, Geurst80, Purcell81} on a variational 
formulation of
the Landau two-fluid equations, with little or no reference to an underlying
Bose
condensate. Indeed, the kind of variational formulation we use was
originally inspired by the goal of giving a clear derivation of the
two-fluid equations for superfluid helium in the absence of a more microscopic
theory based on BEC.  In contrast, our interest in such a variational
formulation lies in its usefulness in finding explicit solutions of the
linearized two-fluid equations corresponding to collective modes in non-uniform
trapped superfluid gases.  However, in Appendix A, we briefly review
some
subtle aspects of the variational formulation of the two-fluid equations.  

Our discussion of the two-fluid hydrodynamic equations using a variational
approach is a natural extension of the approach recently used to deal
with a pure superfluid in trapped gases at $T=0$~\cite{Zubarev04,
Bulgac05, Stringari05}.  In Section VII, we consider the $T=0$ limit of our
formalism to make contact with this ``quantum hydrodynamic" theory, originally
pioneered by Pitaevskii and Stringari~\cite{StringariPit98} to extend the
mean-field Gross-Pitaevskii theory of collective modes~\cite{Stringari96}.  A 
hydrodynamic description means that a few local variables are sufficient to
describe the dynamics.  While such a description requires rapid collisions and
local equilibrium to be valid in the case of a normal fluid, a hydrodynamic
description of a superfluid is always correct.  Thus it is no suprise that the
$T \rightarrow 0$ limit of two-fluid hydrodynamics gives the correct quantum
description of the pure superfluid (see discussion on p. 170 of
Ref.~\cite{Pethickbook}).   

The discussion in Sections II-IV is fairly technical.  Readers primarily
interested
in the final results can immediately go to Section V and the examples in
Section VI.  Detailed numerical predictions for the hydrodynamic mode
frequencies at finite $T$ based on our formal expressions will be presented in
future papers.  

\section{Landau-Khalatnikov Two-Fluid Thermodynamics for a Trapped Gas}
We note that since two-fluid hydrodynamics describes a system in local
equilibrium, all thermodynamic quantities we discuss are functions of position
and time.  This dependence will usually be left implicit.  Even in static
equilibrium in the presence of
a trapping potential, most thermodynamic quantities will still be position
dependent.  

Our variational principle will make use of a 
Lagrangian density of the form \bea
 \mathcal{L} = T - U, \eea where $T$ is the kinetic energy density of
the fluid and $U$ is the internal energy density.  Thus, we need to formulate
the
thermodynamics in terms of the internal energy density $U$, rather than the
total energy $E = T + U$, as is normally done in two-fluid hydrodynamics
(see, for example, p. 521 in Ref.~\cite{LL}). Our approach
will be to obtain $U$ from the total energy density
used by Landau and Khalatnikov (LK) to derive the Landau two-fluid (LTF)
equations~\cite{Khalatnikov}. 
For a trapped two-fluid system, we define the total energy density $E$ as
the sum of the kinetic and potential energy densities,
\bea E = \frac{1}{2}\rho_s \bv_s^2 + \frac{1}{2}\rho_n \bv_n^2 + U + \rho
U_{\text{ext}}, \label{Etot} \eea where $\rho_n$ and $\rho_s$ are the
normal fluid and superfluid densities, $\bv_n$ and $\bv_s$ are the
normal fluid and superfluid velocities, and $\rho = \rho_s + \rho_n$ is the
total density.  Also, \bea U_{\text{ext}}(\br) =
\frac{1}{2}(\omega_x^2 x^2 +
\omega_y^2 y^2 +
\omega_z^2 z^2)\label{trap}\eea is the harmonic trapping potential
\textit{divided
by
the particle
mass}, as follows from our use of the mass density $\rho$, rather than the
number density $n$.  

Following LK, the
total energy density 
$E_0(\br,t)$ as measured in a frame of reference moving with the local
superfluid velocity is 
related to $E$ by
\bea E_0 = E
- \frac{1}{2}\rho \bv_s^2- \bv_s\cdot[\rho_n(\bv_n - \bv_s)].\label{E_0}\eea 
Using the LK identity for local equilibrium \bea dE_0 = T
dS + \mu d\rho + (\bv_n - \bv_s)\cdot
d\left[\rho_n\left(\bv_n - \bv_s\right)\right], \label{dE0} \eea with
Eqs.~(\ref{Etot}) and (\ref{E_0}), we obtain
the following identity for the internal energy density~\cite{note1}:
\bea dU = (\mu - U_{\text{ext}}) d\rho + T dS + \frac{1}{2}(\bv_n - \bv_s)^2
d\rho_n.
\label{dU}\eea  In similar fashion, the LK expression for the
pressure $P = \partial (E_0V)/\partial V$ including the effects of a trapping
potential can be recast in terms of the
internal energy density, giving \bea P = -U -
\rho U_{\text{ext}} + TS + \mu\rho + \frac{1}{2}\rho_n (\bv_n - \bv_s)^2.
\label{Pressure}\eea  The two definitions given by Eqs.~(\ref{dU}) and
(\ref{Pressure}) can be combined to give \bea \rho d\mu = dP -S dT -
\rho_n(\bv_n - \bv_s)\cdot d(\bv_n - \bv_s). \label{dmu}\eea The
three thermodynamic identities given by Eqs.~(\ref{dU}), (\ref{Pressure}), and
(\ref{dmu})
define all thermodynamic properties that we will require in developing our
variational approach.  We note that Eq.~(\ref{dU}) implies that
\bea \mu = \left(\frac{\partial U}{\partial\rho}\right)_{S,\rho_n}\!\! +
\;U_{\text{ext}}. \label{mudef0}\eea

\section{Zilsel's Hydrodynamics}

In both classical and quantum mechanics, one can obtain dynamical
equations of motion by equating to zero the variation of some action, usually
expressed in the form $A
= \int dt(T - U)$, where $T$ and $U$ are the kinetic and potential energies,
respectively.  To derive the LTF equations, we take an action of the
form~\cite{Zilsel50}
\bea
A &=&\int d^{4}x\;\Big[\frac{1}{2}(\rho - \rho_n)\bv_s^2 \nonumber\\&&+
\frac{1}{2}\rho_n \bv_n^2 -
U(\rho, \rho_n, S) - \rho U_{\text{ext}}\Big]. \label{S0}
\eea Here
$d^{4}x = d^{3}r\;dt$, while $U$ is the internal energy defined by
Eq.~(\ref{dU}) and $U_{\text{ext}}$ is the harmonic trapping potential given
by Eq.~(\ref{trap}).   An action similar to Eq.~(\ref{S0}) was originally
used by Zilsel in 1950 to derive the LTF equations.  In this section, we
largely follow Zilsel's approach, with some differences noted in Appendix A.

In
taking
the variation of the action in Eq.~(\ref{S0}), the variables $\bv_n, \bv_s,
\rho, \rho_n$, and
$S$ will be treated as independent so
that, for instance, \bea \frac{\delta U}{\delta \rho} \equiv
\left(\frac{\partial
U}{\partial \rho}\right)_{S,\rho_n}.\eea 

Two
important conservation laws not incorporated into the action given
by Eq.~(\ref{S0}) are the conservation of mass
and entropy, described by:
\bea \frac{\partial \rho}{\partial t} + \bnab\cdot\left[(\rho - \rho_n)
\bv_s + \rho_n\bv_n\right]= 0, \label{rhocont}\\
\frac{\partial S}{\partial t} + \bnab\cdot(S \bv_n) = 0.
\label{scont} \eea The variation
of the action in Eq.~(\ref{S0}) must be taken subject to the constraints
given by Eqs.~(\ref{rhocont}) and (\ref{scont}).  An additional constraint 
not employed by Zilsel describes conservation of circulation in the
normal fluid and allows for the possibility of vorticity.  This constraint takes
the form~\cite{Yourgau, Geurst80, Jackson78} \bea \frac{\partial
S\eta}{\partial t} + \bnab\cdot (S\eta
\bv_n)= 0, \label{circcont}\eea  
where $\eta$ depends on $\br$ and $t$.  Even though the inclusion of this
constraint has no effect on the final form
of the LTF equations, it eliminates the restrictions on the normal fluid
velocity field that were present in Zilsel's original work.  A more complete
discussion
of this point is given in Appendix A.  Following the approach pioneered by
Eckart~\cite{Eckart38}, this constraint, along with entropy and mass
conservation, can
be
incorporated into the variational principle by introducing Lagrange multipliers
$\phi$,
$\alpha$, and $\gamma$ (all dependent on $\br$ and $t$) so that the action
becomes 
\bea
A &=& \int d^{4}x\;\Bigg[\frac{1}{2}(\rho - \rho_n)\bv_s^2 +
\frac{1}{2}\rho_n \bv_n^2 -
U(\rho, \rho_n,S) \nonumber\\&&- \rho U_{\text{ext}} +
\phi\left\{ \frac{\partial \rho}{\partial t} + \bnab\cdot\left[(\rho - \rho_n)
\bv_s + \rho_n\bv_n\right]\right\} \nonumber\\&&+
\alpha\left\{\frac{\partial S}{\partial t} +
\bnab\cdot(S \bv_n)\right\}\nonumber\\
&&+\gamma\left\{\frac{\partial S\eta}{\partial t} +
\bnab\cdot(S\eta \bv_n)\right\}\Bigg]. \label{S0c} \eea  
It is understood that the time and spatial integration in Eq.~(\ref{S0c}) is done
between two fixed points with the fluctuations of the variables appearing in
the action vanishing at both.  

Setting the
variation of the action given by Eq.~(\ref{S0c}) with respect to $\rho, S, \eta,
\bv_s$,
and
$\bv_n$ equal to zero, and making use of the thermodynamic identities implied
by Eq.~(\ref{dU}), we obtain the following relations:
\bea \frac{\delta A}{\delta\rho} &=& \frac{1}{2}\bv_s^2 -\frac{\partial
\phi}{\partial t}
- \bv_s\cdot\bnab\phi - \mu =
0,\label{deltarho}\eea
\bea \frac{\delta A}{\delta S} =-\frac{\partial
\alpha}{\partial t} -
\bv_n\cdot\bnab\alpha - T -\eta\left(\frac{\partial \gamma}{\partial t} +
\bv_n\cdot\bnab\gamma\right) = 0,\nonumber\\ \label{deltaS}\eea
\bea \frac{\delta A}{\delta\eta}  = -S\left(\frac{\partial \gamma}{\partial t} +
\bv
_n\cdot\bnab\gamma\right)= 0,
\label{deltalamb}\eea
\bea \frac{\delta A}{\delta\bv_s} = \rho_s\left(\bv_s - \bnab\phi\right) = \b0,
\label{deltavs}\eea and
\bea \frac{\delta A}{\delta\bv_n} = \rho_n\left(\bv_n -
\bnab\phi\right) - S\left(\bnab\alpha + \eta\bnab\gamma\right) = \b0.
\label{deltavn}\eea  Taking the variation of the
action with respect to $\rho_n$ and using Eq.~(\ref{deltavs}), one recovers a
thermodynamic identity already known from Eq.~(\ref{dU}): \bea
\left(\frac{\partial U}{\partial \rho_n}\right)_{S, \rho} =
\frac{1}{2}(\bv_n - \bv_s)^2. \label{dUdrhon} \eea Additionally, taking the
variation of the action with
respect to $\phi, \alpha$, and $\gamma$ recovers the continuity equation given
by Eq.~(\ref{rhocont}), as well as the entropy and circulation conservation 
laws given by Eqs.~(\ref{scont}) and (\ref{circcont}). 

Equations~(\ref{deltarho})-(\ref{deltavn}) can be rearranged to yield the
useful
expressions
\bea \bv_s = \bnab\phi, \label{vs} \eea
\bea  \frac{\rho_n}{S}\left(\bv_n -
\bv_s\right) = \bnab \alpha  +\eta\bnab\gamma,
\label{nabphi}\eea
\bea \frac{\partial \alpha}{\partial t} + \eta\frac{\partial
\gamma}{\partial t}= -T -
\frac{\rho_n}{S}\bv_n\cdot\left(\bv_n - \bv_s\right),
\label{dphin} \eea
\bea \frac{\partial \gamma}{\partial t} +
\bv_n\cdot\bnab\gamma=0, 
\label{gammacont}\eea
and
\bea 
\frac{\partial \phi}{\partial t} &=& -\left(\mu  
+ \frac{1}{2}\bv_s^2\right).  \label{alpha}\eea  Equation~(\ref{vs})
shows that the superfluid velocity is irrotational.  Furthermore, comparing
Eqs.~(\ref{scont}) and (\ref{circcont}), we find
\bea \frac{\partial \eta}{\partial t} +
\bv_n\cdot\bnab\eta=0.  
\label{betacont}\eea
Another relationship that will be helpful in deriving the equation of motion
for the normal fluid, found from Eqs.~(\ref{nabphi}), (\ref{gammacont}), and
(\ref{betacont}), is
\bea \frac{\partial \eta}{\partial t}\bnab \gamma - \frac{\partial
\gamma}{\partial t}\bnab
\eta = \bv_n \times \left(\bnab\times\left[\frac{\rho_n}{S}(\bv_n
- \bv_s)\right]\right). \label{useful0}\eea

Taking the time derivative of Eq.~(\ref{vs}) and the gradient of
Eq.~(\ref{alpha}),
we find the following equation for the superfluid velocity:
\bea \frac{\partial\bv_s}{\partial t} 
=-\bnab\left(\mu  
+ \frac{1}{2}\bv_s^2\right). \label{vs0}\eea
Rearranging Eq.~(\ref{Pressure}) to obtain an equation for $\mu$, and
making use of the result
\bea \lefteqn{\bnab U=}\nonumber\\&& \left(\frac{\partial U}{\partial
\rho}\right)_{\rho_n,
S}\!\bnab\rho +\;\left(\frac{\partial U}{\partial
\rho_n}\right)_{\rho,
S}\!\bnab\rho_n+\left(\frac{\partial U}{\partial
S}\right)_{\rho,
\rho_n}\!\!\bnab S,\nonumber\\
&=&(\mu - U_{\text{ext}})\bnab\rho + \frac{1}{2}\left(\bv_n -
\bv_s\right)^2\bnab \rho_n +
T\bnab S, \label{gradU} \eea
we obtain the expression \bea \bnab\mu &=& \frac{1}{\rho}\bnab P -\frac{S}{\rho}
\bnab T \nonumber\\&+& \bnab
U_{\text{ext}} - \frac{1}{2}\frac{\rho_n}{\rho}\bnab\left[\left(\bv_n -
\bv_s\right)^2\right]. \label{gradmu} \eea Thus,
Eq.~(\ref{vs0}) for the
superfluid velocity can be rewritten as
\bea \lefteqn{\left(\frac{\partial}{\partial t} + \bv_s\cdot\bnab\right)\bv_s
=\frac{S}{\rho}\bnab T}\nonumber\\
&&
- \frac{1}{\rho}\bnab P -\bnab U_{\text{ext}} +
\frac{1}{2}\frac{\rho_n}{\rho}\bnab\left[\left(\bv_n -
\bv_s\right)^2\right].\label{vst}\eea

To determine the analogous equation for the velocity of the normal fluid, we
take the
time-derivative of Eq.~(\ref{nabphi}) and the gradient of Eq.~(\ref{dphin}). 
Then, using
Eqs.~(\ref{rhocont}), (\ref{scont}),
(\ref{circcont}), (\ref{nabphi}), and (\ref{useful0}),  as well as the identity
$\bv\cdot\bnab\bv = \;\;\;\;\;(1/2)\bnab (\bv^2) - \bv \times (\bnab \times
\bv)$,
after some labourious
algebra, we find
\bea \lefteqn{\left(\frac{\partial}{\partial t} + \bv_n\cdot \bnab\right)\bv_n
=
-\frac{\rho_s}{\rho_n}\frac{S}{\rho} \bnab T }\nonumber\\&&- \frac{1}{\rho}\bnab
P -
\bnab
U_{\text{ext}} - \frac{1}{2}\frac{\rho_s}{\rho}\bnab\left[\left(\bv_n -
\bv_s\right)^2\right]\nonumber\\&& - \frac{\Gamma}{\rho_n}\left(\bv_n -
\bv_s\right),
\label{vnt} \eea
where the ``source function" $\Gamma$ is defined by \bea \Gamma
\equiv \frac{\partial
\rho_n}{\partial t} + \bnab\cdot(\rho_n \bv_n). \label{rhoncont}
\eea 
From the continuity equation given by Eq.~(\ref{rhocont}), Eq.~(\ref{rhoncont})
implies that
\bea \frac{\partial
\rho_s}{\partial t} + \bnab\cdot(\rho_s \bv_s) = -\Gamma. \label{rhoscont}
\eea 
Combining Eqs.~(\ref{vst})-(\ref{rhoscont})
to obtain an
equation of motion for the
current, $\bj = \rho_s\bv_s + \rho_n\bv_n$, the source terms of the two 
components cancel and we find  \bea
\frac{\partial \bj}{\partial t} &=& -\bnab P - \rho\bnab U_{\text{ext}} -
\rho_s \bv_s\cdot \bnab \bv_s - \rho_n\bv_n\cdot\bnab\bv_n\nonumber\\
&&-\bv_s \bnab\cdot (\rho_s \bv_s) - \bv_n \bnab\cdot (\rho_n \bv_n). \label{jt}
\eea  More familiarly, in component form, this equation can be written
as~\cite{Khalatnikov}
\bea \frac{\partial j_i}{\partial t}
&=&-\frac{\partial}{\partial x_j}\left[P\delta_{ij} +
\rho_s v_{si}v_{sj} + \rho_n v_{ni}v_{nj}\right]\nonumber\\
&&-\rho\frac{\partial U_{\text{ext}}}{\partial x_i},\label{jt2}\eea
where the index $j$ is summed over.

Taken together, Eqs.~(\ref{rhocont}), (\ref{scont}),
(\ref{vs0}), and (\ref{jt2}) constitute the Landau two-fluid equations in the
non-dissipative limit,
generalized to include a static external potential. 

In Ref.~\cite{Zilsel50}, Zilsel claims that the source term appearing in
Eq.~(\ref{vnt})
does not appear in the LTF equations of Landau~\cite{Landau41}.
However, Landau only derived
equations of motion for the superfluid velocity, as well as the local
current density.  These are identical to
Eqs.~(\ref{vs0}) and (\ref{jt2}), apart from the external potential we are
including.  If one
works
backwards from these two equations to derive an equation for the normal 
fluid velocity, \textit{the source term in Eq.~(\ref{vnt}) does appear}.  Thus,
the
LTF equations do
include a term that contains $\Gamma$.  We also note that the appearance of
$\Gamma$ in the
equation of motion for the normal fluid velocity is consistent with the
results derived by ZNG from a microscopic model (for further discussion, see
Appendix B).

Even though our variational method has been developed so that we do not have
to solve the linearized LTF equations, to facilitate comparison with that
approach we derive those equations here. 
For the case
where $\bv_{s0} = \bv_{n0} = \b0$, the linearized LTF equations for the
velocity fields are
\bea  \frac{\partial \bv_s}{\partial t} &=& \frac{S_0}{\rho_0}\bnab \delta T -
\frac{1}{\rho_0}\bnab\delta P + \frac{\bnab P_0}{\rho_0^2}\delta
\rho\nonumber\\
&=&-\bnab\delta\mu
\label{vstl} \eea and \bea  \frac{\partial \bv_n}{\partial t} &=&
-\frac{\rho_{s0}}{\rho_{n0}}\frac{S_0}{\rho_0}\bnab \delta T -
\frac{1}{\rho_0}\bnab\delta P + \frac{\bnab P_0}{\rho_0^2}\delta
\rho\nonumber\\
&=&-\bnab\delta\mu - \frac{S_0}{\rho_{n0}}\bnab\delta T. 
\label{vntl} \eea  Here we have made use of the fact that $\Gamma$ is
zero in equilibrium~\cite{ZNGJLTP}.  The last term in Eq.~(\ref{vnt}) is second
order in the fluctuations and thus 
vanishes in a
linearized theory.  We have also made use of the fact that in equilibrium 
$\bnab T_0 = \b0$.  Furthermore, in equilibrium we also have $\bnab
\mu_0 = \b0$ so that for $\bv_{n0} = \bv_{s0} = \b0$, Eq.~(\ref{gradmu})
yields
the well-known relation \bea \bnab U_{\text{ext}} = -\frac{1}{\rho_0}\bnab
P_0.
\label{useful} \eea Using this, the linearized hydrodynamic velocity equations
reduce to
\bea  \frac{\partial \bv_s}{\partial t} = \frac{S_0}{\rho_0}\bnab \delta T -
\frac{1}{\rho_0}\bnab\delta P - \frac{\bnab U_{\text{ext}}}{\rho_0}\delta
\rho,
\label{vstl2} \eea \bea  \frac{\partial \bv_n}{\partial t} =
-\frac{\rho_{s0}}{\rho_{n0}}\frac{S_0}{\rho_0}\bnab \delta T -
\frac{1}{\rho_0}\bnab\delta P- \frac{\bnab U_{\text{ext}}}{\rho_0}\delta
\rho.   
\label{vntl2} \eea Combining these, we obtain
\bea \frac{\partial \delta \bj}{\partial t} = -\bnab\delta P - \delta\rho
\bnab U_{\text{ext}}. \label{jtl} \eea
\section{Action for Linearized Landau Two-Fluid Hydrodynamics}

As was shown in the previous section, variation
of the action given by Eq.~(\ref{S0c}) with respect to $\rho$, $S$,
$\eta$, $\bv_s$, $\bv_n$, $\phi$, $\alpha$, and $\gamma$ leads to the
non-dissipative LTF
hydrodynamic
equations~\cite{Zilsel50}. To determine the low-energy collective modes of
this
system given by the solutions of the linearized hydrodynamic
equations, we
Taylor-expand the action about the equilibrium values of these variables. 
In principle, one could then take the variation of the resulting action with
respect to
the fluctuations of its variables, i.e., $\delta\rho$, $\delta S$,
$\delta\eta$, $\delta\bv_s$, $\delta\bv_n$, $\delta\phi$, $\delta\alpha
$, and $\delta\gamma$, to obtain the linearized hydrodynamic equations.  These
could then be solved
to find the
collective mode
frequencies.  In
practice, rather than dealing with the Lagrange multipliers $\phi$, $\alpha$,
and $\gamma$ in Eq.~(\ref{S0c}), it is more convenient to
introduce displacement fields for the two velocities.  This allows one to
incorporate the
conservation laws directly into expressions for $\delta\rho$ and $\delta
S$, thereby eliminating the need for Lagrange multipliers.   We will
employ a simplified Rayleigh-Ritz method in
conjunction with our variational approach to obtain estimates of the collective
mode frequencies by using physically reasonable ansatzes for the displacement
fields of the superfluid and normal fluid.

Recalling that the
conservation of
circulation constraint does not affect the form of the equations resulting from
the variational principle, we will omit this term in the
action for sake of clarity.  However,  if we were interested in
collective oscillations of the normal fluid with non-zero circulation, we would
need to
incorporate this
term and allow for fluctuations of $\eta$ and
$\gamma$.  

Considering fluctuations of the action about equilibrium we set
$\rho
= \rho_0 + \delta \rho$, $\rho_n
= \rho_{n0} + \delta \rho_n$, $S = S_0 +\delta S$,
$\phi= \phi_0 + \delta\phi$, $\bv_n = \bv_{n0} + \delta\bv_n$, $\bv_s = \bv_{s0}
+ \delta\bv_s$, and expand the action given by Eq.~(\ref{S0c}) up to
quadratic
order in these fluctuations.  As in the above, we assume that the
two
fluid components are stationary in equilibrium, so that $\delta\bv_n=\bv_n$ and
$\delta\bv_s=\bv_s$.   Writing $A = A^{(0)} +
A^{(1)} + A^{(2)}
+
\cdots$, the contribution to the action
from terms which are quadratic
in the fluctuations is 
\bea
\lefteqn{A^{(2)} =}\nonumber\\&&\int
d^{4}x\;\Bigg[\frac{1}{2}\rho_{s0}\bv_s^2 +
\frac{1}{2}\rho_{n0} \bv_n^2 \;-\;
\frac{1}{2}\left(\frac{\partial \mu}{\partial
\rho}\right)_{S,\rho_n}\!\!\!(\delta\rho)^2 \nonumber\\&& -
\;\left(\frac{\partial
T}{\partial\rho}\right)_{S,\rho_n}\!\!\! \delta S \delta\rho \;-\;\frac{1}{2}
\left(\frac{\partial T}{\partial
S}\right)_{\rho,\rho_n}\!\!\!(\delta S)^2 \nonumber\\
&&+\delta\phi\left\{\frac{\partial \delta \rho}{\partial t} +
\bnab\cdot\left(\rho_{s0}\bv_s +
\rho_{n0}\bv_n\right)\right\} \nonumber\\
&&+\phi_0\left\{\bnab\cdot\Big[(\delta\rho-\delta\rho_n)\bv_s +
\delta\rho_n\bv_n\Big]\right\}\nonumber\\
&&+\delta\alpha\left\{\frac{\partial \delta S}{\partial t} +
\bnab\cdot\left(S_{0}\bv_n\right)\right\} \nonumber\\&&+
\alpha_0\left\{\bnab\cdot(\delta S \bv_n)\right\}\Bigg], \label{S0c2a}\eea where
we have made
use of Eq.~(\ref{dU}) to rewrite the coefficients in terms of thermodynamic
derivatives. Note that there is an alternative expression for the 
coefficient multiplying
$\delta S\delta\rho$, given by the Maxwell relation \bea
(\partial
T/\partial\rho)_{S} = (\partial
\mu/\partial S)_{\rho}.\label{maxwell}\eea
Contributions to the fluctuations of the internal energy arising
from fluctuations of $\rho_n$ have vanished
since, as can be seen from Eq.~(\ref{dUdrhon}), the
equilibrium
value $(\partial U/\partial \rho_n)_{S,\rho}$ is zero when $\bv_{n0} =
\bv_{s0}$.
We do not consider the contributions which are linear
in
fluctuations since we know from the stationary action
principle
that $A^{(1)} = 0$.  

Since $\bv_{s0} = \b0$, Eq.~(\ref{vs}) gives us $\bnab\phi_0 = \b0$. 
Also, with $\bv_{n0} = \bv_{s0}=\b0$,
from Eq.~(\ref{nabphi}) we find $\bnab\alpha_0 =\b0$ since $\eta$ is an
independent and
arbitrary variable.  With 
these relations, having integrated by parts the terms in Eq.~(\ref{S0c2a})
which contain $\phi_0$ and $\alpha_0$, the quadratic action simplifies to
\bea
\lefteqn{A^{(2)} =}\nonumber\\&&\int
d^{4}x\;\Bigg\{\frac{1}{2}\rho_{s0}\bv_s^2 +
\frac{1}{2}\rho_{n0} \bv_n^2 \;-\;
\frac{1}{2}\left(\frac{\partial \mu}{\partial
\rho}\right)_{S,\rho_n}\!\!\!(\delta\rho)^2 \nonumber\\&& -
\;\left(\frac{\partial
T}{\partial\rho}\right)_{S,\rho_n}\!\!\! \delta S \delta\rho \;-\;\frac{1}{2}
\left(\frac{\partial T}{\partial
S}\right)_{\rho,\rho_n}\!\!\!(\delta S)^2 \nonumber\\ 
&&+\delta\phi\left[\frac{\partial \delta \rho}{\partial t} +
\bnab\cdot\left(\rho_{s0}\bv_s +
\rho_{n0}\bv_n\right)\right] \nonumber\\
&&+\delta\alpha\left[\frac{\partial \delta S}{\partial t} 
+\bnab\cdot(S_0\bv_n)\right]\Bigg\}. \label{S0c2}\eea
In the present form, the variational principle has been reduced to
equating to zero the variation of the action given by the first two lines in
Eq.~(\ref{S0c2}), subject to the constraints given by the linearized
continuity and entropy conservation equations,
\bea \frac{\partial \delta \rho}{\partial t} +
\bnab\cdot\left(\rho_{s0}\bv_s +
\rho_{n0}\bv_n\right)=0 \label{rhocontl}\eea
and
\bea \frac{\partial \delta S}{\partial t} 
+\bnab\cdot\left(S_0\bv_n\right)=0.\label{scontl}\eea
For our purposes, it is more
convenient to incorporate these two constraints into expressions for
$\delta\rho$ and $\delta S$ than to use the Lagrange multipliers $\delta
\phi$ and $\delta\alpha$. Introducing the displacement fields \bea
\bv_s(\br,t) \equiv \frac{\partial \bu_s(\br,t)}{\partial t}, \; \; \;
\bv_n(\br,t) \equiv \frac{\partial \bu_n(\br,t)}{\partial t}, \label{fields}
\eea the linearized continuity and entropy conservation equations can be
expressed in terms of these 
fields as\bea \delta
\rho(\br,t) = -\bnab\cdot\left[\rho_{s0}(\br)\bu_s(\br,t) +
\rho_{n0}(\br)\bu_n(\br,t)\right] \label{rhocontdf} \eea and
\bea\delta S(\br, t) &=&
-\bnab \cdot \left[S_0(\br)\bu_n(\br,t)\right].\label{scontdf} \eea Removing
the terms
that contain the Lagrange
multipliers from Eq.~(\ref{S0c2}) and substituting these expressions for
$\delta\rho$ and $\delta S$, we find our central result for the action
that is second order in the fluctuations $\bu_s$ and $\bu_n$,
\bea
\lefteqn{A^{(2)} =}\nonumber\\&&
\int d^4x\;\Bigg\{\frac{1}{2}\rho_{s0}\dot{\bu}_s^2 +
\frac{1}{2}\rho_{n0}\dot{\bu}_n^2 \nonumber\\&&-
\frac{1}{2}\left(\frac{\partial
\mu}{\partial\rho}\right)_{S,\rho_n}\left[\bnab\cdot\left(\rho_{s0}\bu_s
+\rho_{n0}\bu_n\right)\right]^2
\nonumber\\&&-\left(\frac{\partial T}{\partial
\rho}\right)_{S,\rho_n}\left[\bnab\cdot\left(S_0\bu_n\right)\right]\left[\bnab
\cdot\left(\rho_{s0}\bu_s
+\rho_{n0}\bu_n\right)\right]
\nonumber\\&&-
\frac{1}{2}\left(\frac{\partial T}{\partial
S}\right)_{\rho,\rho_n}\left[\bnab\cdot\left(S_0\bu_n\right)\right]^2\Bigg\}.
\label{S0c3} \eea   

To make our expressions as compact as possible, in what follows we will drop the
$\rho_n$ term constraining the thermodynamic derivatives in Eq.~(\ref{S0c3}).  
However, it should be remembered that in evaluating the various
local thermodynamic derivatives, $\rho_n$ is always fixed. 
\section{hydrodynamic modes describing fluctuations}
The
fluctuating quantities  $\delta\phi$, $\delta\alpha$, $\delta S$,
$\delta\rho$, $\bv_s$, and $\bv_n$ appearing in Eq.~(\ref{S0c2}),
have been replaced in Eq.~({\ref{S0c3}) by the two displacement fields $\bu_n$
and $\bu_s$. Thus, the variational principle has been reduced to taking the
variation of
the quadratic action given by Eq.~(\ref{S0c3}) with respect to only
$\bu_n$ and $\bu_s$.  The linearized LTF hydrodynamic equations and hence,
low-energy collective modes of the
system, 
are completely determined by the variational equations \bea \frac{\delta
A^{(2)}}{\delta \bu_s(\br,t)} = {\b0},\;\;\;\frac{\delta
A^{(2)}}{\delta \bu_n(\br,t)} = {\b0}.\label{var-1}\eea
Using Eq.~(\ref{S0c3}) as well as the identities given by
Eqs.~(\ref{rhocontdf}) and (\ref{scontdf}), Eq.~(\ref{var-1}) gives
\bea \ddot{\bu}_s &=&
-\bnab\Big[\delta \rho\left(\frac{\partial
\mu}{\partial\rho}\right)_{S}
+\delta S\left(\frac{\partial
\mu}{\partial S}\right)_{\rho}\Big]\label{usddt}\eea 
and
\bea \ddot{\bu}_n&=&
-\bnab\Big[\delta\rho\left(\frac{\partial
\mu}{\partial\rho}\right)_{S}+\delta S
\left(\frac{\partial \mu}{\partial S}\right)_{\rho}\Big]\nonumber\\&-&
\frac{S_0}{\rho_{n0}}\bnab\Big[\delta\rho
\left(\frac{\partial
T}{\partial\rho}\right)_{S}+\delta S
\left(\frac{\partial
T}{\partial S}\right)_{\rho}\Big]\label{unddt},
\eea where we have used the Maxwell relation given in Eq.~(\ref{maxwell}).
Taking $\delta F = \delta S(\partial F/\partial S)_{\rho} +
\delta\rho(\partial F/\partial \rho)_{S}$, where $F$ denotes either
one
of $\mu$ or $T$, with the displacement fields defined by Eqs.~(\ref{fields}),
it is apparent that Eqs.~(\ref{usddt}) and (\ref{unddt}) are equivalent to the
linearized
two-fluid hydrodynamic equations given by Eqs.~(\ref{vstl}) and
(\ref{vntl}).  This verifies that the variational principle given by Eqs.~(\ref
{S0c3}) and (\ref{var-1}) is the correct one.   

Solutions of Eq.~(\ref{var-1}) corresponding to collective modes will have
harmonic time-dependence of the form
\bea \bu_s(\br,t) = \bu_s(\br)\cos(\omega t),\;\;\bu_n(\br,t)
=\bu_n(\br)\cos(\omega t). \label{harmonic} \eea 
Because these are exact solutions of the variational equations, we can insert
these expressions into the action directly.  Substituting Eq.~(\ref{harmonic})
into Eq.~(\ref{S0c3}) and performing the time-integration, we obtain, apart from
an irrelevant constant factor, the following Lagrangian:
\bea
L^{(2)} = K[\bu_s, \bu_n]\omega^2 - U[\bu_s, \bu_n], 
\label{L0c3} \eea where
\bea
K[\bu_s, \bu_n] = 
\frac{1}{2}\int d^3r\;\left\{\rho_{s0}\bu_s^2 +
\rho_{n0}\bu_n^2\right\} \label{K}\eea 
and
\bea \lefteqn{U[\bu_s, \bu_n] =}\nonumber\\&& \frac{1}{2}\int d^3r\;\Bigg\{
\left(\frac{\partial
\mu}{\partial\rho}\right)_{S }\left[\bnab\cdot\left(\rho_{s0}\bu_s
+\rho_{n0}\bu_n\right)\right]^2
\nonumber\\&&+2\left(\frac{\partial T}{\partial
\rho}\right)_{S }\left[\bnab\cdot\left(S_0\bu_n\right)\right]\left[\bnab
\cdot\left(\rho_{s0}\bu_s
+\rho_{n0}\bu_n\right)\right]
\nonumber\\&&+
\left(\frac{\partial T}{\partial
S}\right)_{\rho }\left[\bnab\cdot\left(S_0\bu_n\right)\right]^2\Bigg\}.
\label{U} \eea 
Since the fields $\bu_s(\br)$ and $\bu_n(\br)$ are no longer time-dependent, it
suffices to consider variations of the Lagrangian given by Eq.~(\ref{L0c3}), and
the variational equations now become
\bea \frac{\delta L^{(2)}}{\delta \bu_s(\br)} = \b0,\;\;\;
\frac{\delta L^{(2)}}{\delta \bu_n(\br)} = \b0.
\label{var1}\eea

Solving Eq.~(\ref{var1}) is still equivalent to solving the
linearized hydrodynamic equations, with the only simplification being that we
have assumed harmonic time-dependence for the solutions.
Fortunately, one can easily obtain approximate expressions for the collective
mode frequencies within a
variational approach. 
Following Ref.~\cite{ZNGJLTP}, we use a simplified Rayleigh-Ritz method and
make an ansatz for each Cartesian component of the displacement fields of the
form
\bea u_{si} (\br)=  a_{si}f_{i}(\br),\;\;\;
u_{ni} (\br)&=& a_{ni}g_{i}(\br), \label{RR} \eea
where the constant coefficients $a_{si}$ and $a_{ni}$ are variational
parameters.  Substituting this ansatz into Eq.~(\ref{L0c3}), and equating to
zero the variation of the resulting expression with respect to these
parameters, we have the six variational equations \bea \frac{\delta
L^{(2)}}{\delta a_{si}} = 0, \;\;\;
\frac{\delta L^{(2)}}{\delta a_{ni}} = 0.  \label{var2}\eea
In practice, the symmetry of the problem usually allows us to reduce the
number of equations.  From these
equations, one obtains a rigorous \textit {upper bound}
for the collective mode frequencies $\omega$~\cite{RR}. 

A more precise Rayleigh-Ritz method would require us to expand each component of
the displacement fields in terms of $n \;(>1)$ elements of some complete set
of functions.   As $n$ is
increased, solving the resulting $6n$ variational equations, one iteratively
generates successively better approximations for the collective mode
frequencies.  For
harmonically-confined gases, however, there exist
simple trial functions for $f_i(\br)$ and $g_{i}(\br)$ which are
sufficiently close to the exact solutions that excellent results
are obtained by considering only a single expansion term as in
Eq.~(\ref{RR})~\cite{ZNGJLTP}. 
Our choices of ansatzes for the displacement fields at finite
temperatures will be guided by the known exact solutions at
$T=0$~\cite{Stringari96} and $T>T_c$ (for bosons, see
Ref.~\cite{GriffinStringari}; for fermions, see Ref.~\cite{Bruun99})
for modes of experimental interest, such as the dipole and breathing modes.  

Using the ansatz given by Eq.~(\ref{RR}), the Lagrangian given by
Eqs.~(\ref{L0c3})-(\ref{U}) describes the dynamics of
a pair of coupled harmonic
oscillators, with $a_{si}$ and $a_{ni}$ representing the displacements of the
two oscillators from equilibrium.  The effective spring constants are
determined by the equilibrium thermodynamic quantities of the system.  This is a
useful picture to have
when envisioning the low-energy dynamics of the two fluids.  It immediately
implies, for instance, the existence of in-phase as well as out-of-phase
oscillation modes of the two fluids. 
\section{Examples of Collective Modes}
\subsection{Uniform Gas}
As a simple
example of our formalism, we consider first a uniform
gas.  In this case, $\rho_{s0}, \rho_{n0}$, and the equilibrium thermodynamic
derivatives appearing in Eqs.~(\ref{K}) and (\ref{U}) are all independent of
position. 
Anticipating sound modes with a dispersion relation of the form $\omega = uk$, where $u$ is the
sound speed, we use the plane-wave ansatz \bea \bu_s(\br) &=&
\hat{\bx}\;\mathcal{N}a_s\cos(\omega x/u),\nonumber\\
 \label{ansatzh}\\
\bu_n(\br)
&=&
\hat{\bx}\;\mathcal{N}a_n\cos(\omega x/u),\nonumber
\eea  where $a_s$ and $a_n$ are the variational parameters.  The normalization
constant $\mathcal{N}$ is chosen so that $\int d^{3}r\;\bu_s^2 = a_s^2$. 
Inserting Eq.~(\ref{ansatzh}) into Eqs.~(\ref{K}) and (\ref{U}),
we find 
\bea K[a_s, a_n] = \frac{1}{2}\rho_{s0} a^2_s +  \frac{1}{2}\rho_{n0} a^2_n
 \eea  and
\bea \lefteqn{U[a_s, a_n] = \frac{\omega^2}{u^2}\Bigg\{
\frac{a^2_s}{2}\left[(\rho_{s0})^2\left(\frac{\partial
\mu}{\partial\rho}\right)_{S}\right]}\nonumber \\&
+&a_sa_n\left[\rho_{s0}\rho_{n0}\left(\frac{\partial
\mu}{\partial\rho}\right)_{S}  + S_0\rho_{s0} \left(\frac{\partial T}{\partial
\rho}\right)_{S}\right]\nonumber\\&+&
\frac{a^2_n}{2}\Bigg[(\rho_{n0})^2 \left(\frac{\partial
\mu}{\partial\rho}\right)_{S} + 2S_0\rho_{n0} \left(\frac{\partial T}{\partial
\rho}\right)_{S} \nonumber\\&+& (S_0)^2   \left(\frac{\partial T}{\partial
S}\right)_{\rho}\Bigg]\Bigg\}. \eea 
The trial functions given by Eq.~(\ref{ansatzh}) are, in fact, exact.  Thus, 
the
variational equations given by Eq.~(\ref{var2}), which in the present case take
the form
\bea \frac{\delta L^{(2)}}{\delta a_s} = 0, \;\;\;
\frac{\delta L^{(2)}}{\delta a_n} = 0,  \label{var2h}\eea
will give the exact eigenvalues $\omega$ or, equivalently, the exact sound
speeds $u$.      
Using the above expressions for $K$ and $U$,  Eqs.~(\ref{L0c3}) and
(\ref{var2h}) give the
following quadratic equation for $u^2$:
\bea u^4 &-& u^2\left[\rho_0\left(\frac{\partial \mu}{\partial
\rho}\right)_{S} + 2S_0\left(\frac{\partial T}{\partial \rho}\right)_S + 
\frac{(S_0)^2}{\rho_{n0}}\left(\frac{\partial T}{\partial
S}\right)_{\rho}\right] \nonumber\\&&+
\frac{\rho_{s0}}{\rho_{n0}}(S_0)^2
\frac{\partial(\mu, T)}{\partial(\rho, S)}=0. \label{sss} \eea Here
$\partial(\mu, T)/\partial (\rho, S)$ denotes the Jacobian of the transformation
between $\mu, T$ and $\rho, S$.

In the usual textbook discussions, one works with the entropy density $\sigma =
S/\rho$,
rather than the entropy $S$~\cite{Khalatnikov, Pitaevskiibook, Pethickbook}.  To
express
Eq.~(\ref{sss}) in terms of
$\sigma$, we use the following identities which are found from the
transformation properties of thermodynamic derivatives  (see, for example,
chap. 16 in Ref.~\cite{LLSM}):
\bea
\left(\frac{\partial\mu}{\partial\rho}\right)_{S} =
-2\sigma \left(\frac{\partial T}{\partial \rho}\right)_{\sigma}  +   \frac{\sigma^2}{\rho}\left(\frac{\partial T}{\partial \sigma}\right)_{\rho} +
\frac{1}{\rho}\left(\frac{\partial P}{\partial \rho}\right)_{\sigma}, \eea
\bea
\left(\frac{\partial T}{\partial\rho}\right)_{S} =
\left(\frac{\partial T}{\partial \rho}\right)_{\sigma}  -  
\frac{\sigma}{\rho}\left(\frac{\partial T}{\partial \sigma}\right)_{\rho}, \eea 
\bea \frac{\partial(\mu, T)}{\partial(\rho, S)} =\frac{1}{\rho^2} \left(\frac{\partial
P}{\partial
\rho}\right)_{T}\left(\frac{\partial T}{\partial \sigma}\right)_{\rho},  \eea
and
\bea
\left(\frac{\partial T}{\partial S}\right)_{\rho} =
\frac{1}{\rho}\left(\frac{\partial T}{\partial \sigma}\right)_{\rho} . \eea
With these relations, Eq.~(\ref{sss}) for the sound speed can be rewritten 
as \bea
u^4 &-& u^2\left[\left(\frac{\partial P}{\partial
\rho}\right)_{\sigma} +
\frac{\rho_{s0}}{\rho_{n0}}(\sigma_0)^2\left(\frac{\partial T}{\partial
\sigma}\right)_{\rho}\right] \nonumber\\&&+ \frac{\rho_{s0}}{\rho_{n0}}\sigma_0^2 \left(\frac{\partial
P}{\partial
\rho}\right)_{T}\left(\frac{\partial T}{\partial \sigma}\right)_{\rho}=0
.\label{ss} \eea   This is the standard
equation giving the first and second sound speeds in a
uniform superfluid~\cite{Landau41, Khalatnikov, Pitaevskiibook, Pethickbook}. 

\subsection{Dipole Mode}

We next
consider a dipole mode in a harmonically confined gas.  This mode is
characterized by displacements of the
centre-of-masses of the two fluids along one of the axes of the
harmonic trap, $x_i$.  In this case, we use the following ansatz for the
displacement fields:
\bea \bu_s =
\hat{\bx}_ia_s,\;\;\;\bu_n
=
\hat{\bx}_ia_n, \label{ansatzdip}
\eea
where $a_s$ and $a_n$ describe the displacements of the
centre-of-masses of the two fluids from the trap centre. Substituting
this ansatz into Eqs.~(\ref{K}) and (\ref{U}), we find
\bea
K[a_s, a_n] = 
\frac{1}{2}M_s a^2_s +\frac{1}{2}M_n a^2_n \label{Kd}\eea 
and \bea 
U[a_s, a_n] = \frac{1}{2}k_s a^2_s + \frac{1}{2}k_n a^2_n + \frac{1}{2}k_{sn}
\left(a_s - a_n\right)^2, \label{Ud}\eea
where  $M_s$ and $M_n$ are the 
the masses of the superfluid and normal fluids, respectively, given by \bea M_s
= \int d^{3}r\;\rho_{s0},\;\;\;M_n = \int
d^{3}r\;\rho_{n0}.  \eea The spring constants $k_s, k_n$, and $k_{sn}$ are
given by
\bea k_s &=& \int d^{3}r\;\Bigg\{\left[\left(\frac{\partial
\mu}{\partial
\rho}\right)_{S}\frac{\partial \rho_{0}}{\partial
 x_i} + \left(\frac{\partial \mu}{\partial
S}\right)_{\rho}\frac{\partial S_{0}}{\partial
 x_i}\right]\frac{\partial
\rho_{s0}}{\partial  x_i}\Bigg\}, \nonumber\\ 
 \label{ksd}\eea
\bea k_n &=& \int d^{3}r\;\Bigg\{\left[\left(\frac{\partial
\mu}{\partial
\rho}\right)_{S}\frac{\partial \rho_{0}}{\partial
 x_i} + \left(\frac{\partial \mu}{\partial
S}\right)_{\rho}\frac{\partial S_{0}}{\partial
 x_i}\right]\frac{\partial \rho_{n0}}{\partial  x_i}\nonumber\\
&&+\left[\left(\frac{\partial
T}{\partial
\rho}\right)_{S}\frac{\partial \rho_{0}}{\partial
 x_i} + \left(\frac{\partial T}{\partial
S}\right)_{\rho}\frac{\partial S_{0}}{\partial
 x_i}\right]\frac{\partial S_{0}}{\partial  x_i}\Bigg\}, \label{knd} \eea
and \bea 
\lefteqn{k_{sn} =}\nonumber\\&-&\int d^{3}r\;\Bigg\{\left(\frac{\partial
\mu}{\partial
\rho}\right)_{S}\frac{\partial \rho_{n0}}{\partial
 x_i}\frac{\partial \rho_{s0}}{\partial  x_i} + \left(\frac{\partial T}{\partial
\rho}\right)_{S}\frac{\partial S_{0}}{\partial
 x_i}\frac{\partial \rho_{s0}}{\partial  x_i}\Bigg\}.\nonumber\\
\label{ksndipole}\eea  The variational equation,
equivalent to the expression for the homogeneous gas given by Eq.~(\ref{var2h})
with $K$ and $U$ given by Eqs.~(\ref{Kd}) and (\ref{Ud}), gives the following
secular equation for the dipole frequencies:
\bea \left(
\begin{array}{cc}M_s\omega^2  - k_s - k_{sn}  &
k_{sn}\\ k_{sn} & M_n\omega^2 -k_n - k_{sn}  \end{array}
\right)\left(\begin{array}{c} a_s \\ a_n
\end{array}\right) = 0.\nonumber
\\ \label{secdip}\eea

Applying some thermodynamic identities, the
spring
constants $k_s$ and $k_n$ can be simplified considerably.  Specifically, with
$\bv_{n0}
= \bv_{s0} = \b0$, Eq.~(\ref{dU}) can be used to find useful
expressions for the gradient of various equilibrium thermodynamic
quantities. One
finds, for instance, from the definition $T = (\partial U/\partial S)_{\rho}$, 
\bea \bnab T_0 &=& \left(\frac{\partial^2 U}{\partial
S\partial\rho}\right)\bnab\rho_0 + \left(\frac{\partial^2 U}{\partial
S^2}\right)_{\rho}\bnab
S_0 \nonumber\\
&=&\left(\frac{\partial T}{\partial \rho}\right)_{S}\bnab\rho_0 + 
\left(\frac{\partial T}{\partial S}\right)_{\rho}\bnab S_0 .
\label{bnabT0}\eea
Similarly, using Eq.~(\ref{mudef0}), 
\bea \bnab\mu_0 =\left(\frac{\partial \mu}{\partial \rho}\right)_{S}\bnab\rho_0
+\left(\frac{\partial \mu}{\partial S}\right)_{\rho}\bnab S_0 +\bnab
U_{\text{ext}}, \label{bnabmu}\eea which, it can be shown, is equivalent to
Eq.~(\ref{gradmu})
evaluated at equilibrium, when $\bv_{n0} = \bv_{s0} = \b0$. 
For a harmonic trapping
potential given by
Eq.~(\ref{trap}), noting that $\bnab\mu_0 = \b0$, Eq.~(\ref{bnabmu}) provides
us with the useful expression
\bea\left(\frac{\partial
\mu}{\partial \rho}\right)_S 
\frac{\partial \rho_0}{\partial x_i} + \left(\frac{\partial
\mu}{\partial
S}\right)_{\rho}\frac{\partial S_0}{\partial x_i}&=&-\frac{\partial}{\partial 
x_i}U_{\text{ext}}\nonumber\\&=&
-\omega_i^2x_i, \label{useful3} \eea  where $\omega_i$ is the trap
frequency along the $x_i$-axis.  Also, since $\bnab T_0 = \b0$,
Eq.~(\ref{bnabT0}) gives us
\bea\left(\frac{\partial
T}{\partial \rho}\right)_S 
\frac{\partial \rho_0}{\partial x_i} + \left(\frac{\partial
T}{\partial
S}\right)_{\rho}\frac{\partial S_0}{\partial x_i}=0. \label{dTdx}\eea

Substituting Eqs.~(\ref{useful3}) and (\ref{dTdx}) into Eqs.~(\ref{ksd}) and
(\ref{knd}) and integrating by parts, the expressions for $k_s$ and $k_n$
readily simplify, reducing to $ k_s = \omega_i^2M_s$ and $ k_n =
\omega_i^2M_n$.   With these values, Eq.~(\ref{secdip}) becomes
\bea \left[M_sM_n\left(\omega^2 - \omega_i^2\right) - k_{sn}\left(M_s +
M_n\right)\right]\left(\omega^2 - \omega_i^2\right) = 0. \nonumber\\ \eea
This gives an in-phase dipole mode with frequency \bea \omega = \omega_i
\label{firstsound} \eea and an out-of-phase dipole mode with frequency\bea
\omega^2 =
\omega_i^2 + \frac{M_s + M_n}{M_sM_n}k_{sn}. \label{secondsound} \eea

The result given by Eq.~(\ref{firstsound}) is the expected Kohn mode in a
harmonic trap and corresponds to the solution $a_s = a_n$.  This mode 
is a rigid oscillation of
the superfluid and normal fluid static distributions and as
a result, the
interactions have no effect on the collective mode frequency.  The frequency of
the second mode
given by Eq.~(\ref{secondsound}) does depend on
interactions as these determine the thermodynamic functions appearing in the
$k_{sn}$ spring constant given by Eq.~(\ref{ksndipole}).  This mode
corresponds to the
solution $M_sa_s + M_na_n = 0$.  This is precisely the analogue in a trapped
system of second
sound, since the displacements of the superfluid and
normal fluid have opposite signs, producing an out-of-phase
oscillation of the two components. 

The expression for the frequency of the out-of-phase dipole mode given by
Eq.~(\ref{secondsound}) agrees with that found in
Refs.~\cite{ZNGJLTP} and
\cite{ZNGPRA98}, but with a slightly
different value of $k_{sn}$ in Eq.~(\ref{ksndipole}) as a result of working
with a modified form of two-fluid hydrodynamics (see Appendix B for
further discussion). 

\subsection{Breathing Modes} 
For the breathing modes, we
consider an ansatz of the form 
\bea \bu_s = (a_{s1}x, a_{s2}y, a_{s3}z),\;\;\bu_n = (a_{n1}x, a_{n2}y,
a_{n3}z). \label{monoansatz}\eea

Substituting Eq.~(\ref{monoansatz}) into Eq.~(\ref{L0c3}), we
find
\bea
K[a_s, a_n] = \frac{1}{2}\sum_i\left[M_{si}a_{si}^2 +
M_{ni}a_{ni}^2\right],\label{Km} \eea
and \bea 
U[a_s, a_n] &=&\frac{1}{2}\sum_{ij}\Big[k_{s,ij} a_{si}a_{sj} +
k_{n,ij} a_{ni}a_{nj}\nonumber\\&&\;\;\;\;+ 2k_{sn,ij}a_{si}a_{nj}\Big],
\label{Um}\eea
where the weighted masses $M_i$ are defined by
\bea M_{si} \equiv \int d^{3}r\;\rho_{s0}x_i^2 ,\;\;\;\;M_{ni} \equiv \int
d^{3}r\;\rho_{n0}x_i^2, \label{breathingmasses}\eea
and the
constants $k_{s,ij}, k_{n,ij}$, and $k_{sn,ij}$ (which can be related to 
spring constants) are 
\bea k_{s,ij} &=& \int d^{3}r\;\Bigg\{\left(\frac{\partial
\mu}{\partial
\rho}\right)_{S}\frac{\partial(\rho_{s0}x_i)}{\partial
x_i}\frac{\partial(\rho_{s0}x_j)}{\partial
x_j}\Bigg\}, \label{ksm} \eea
\bea k_{n,ij} &=& \int d^{3}r\;\Bigg\{\left(\frac{\partial
\mu}{\partial
\rho}\right)_{S}\frac{\partial(\rho_{n0}x_i)}{\partial
x_i}\frac{\partial(\rho _{n0}x_j)}{\partial x_j} \nonumber\\&&+
2\left(\frac{\partial T}{\partial
\rho}\right)_{S}\frac{\partial(\rho_{n0}x_i)}{\partial x_i}
\frac{\partial(S_{0}x_j)}{\partial x_j}\nonumber\\
&&+
\left(\frac{\partial T}{\partial
S}\right)_{\rho}\frac{\partial(S_{0}x_i)}{\partial
x_i}\frac{\partial(S_{0}
x_j)}{\partial x_j}\Bigg\}, \label{knm}\eea
and \bea 
\lefteqn{k_{sn,ij} =}\nonumber\\&&\int d^{3}r\;\Bigg\{\left(\frac{\partial
\mu}{\partial
\rho}\right)_{S}\frac{\partial(\rho_{s0}x_i)}{\partial
x_i}\frac{\partial(\rho_{n0}x_j)}{\partial x_j} \nonumber\\&&+
\left(\frac{\partial T}{\partial
\rho}\right)_{S}\frac{\partial(\rho_{s0}x_i)}{\partial
x_i}\frac{\partial(S_0x_j)}{\partial x_j}\Bigg\}.\label{ksnm}\eea 
Since we generally have six variational parameters (one for each Cartesian
component of the two displacement fields), the collective mode frequencies are
found from the six coupled algebraic
equations, $\delta
L^{(2)}/\delta a_{si} = 0,\;\delta L^{(2)}/\delta a_{ni} = 0$.  We find
\bea \lefteqn{M_{si}\omega^2 a_{si} =}\nonumber\\&&
\!\!\frac{1}{2}\sum_j\Big[(k_{s,ij} + k_{s,ji})a_{sj}+ 2k_{sn, ij}a_{nj}\Big],
 \label{monopoles}\eea
\bea \lefteqn{M_{ni}\omega^2 a_{ni}=}\nonumber\\&&
\!\!\frac{1}{2}\sum_j\Big[(k_{n,ij}+ k_{n,ji})a_{nj} + 2k_{sn, ji}a_{sj}\Big].
 \label{monopolen}\eea
In experiments performed on trapped superfluid gases, one usually has
an
axisymmetric trap such that, for instance, $\omega_1 = \omega_2 \equiv
\omega_{\bot},\;\omega_3 =
\omega_z$, and the modes of
interest are the axial and longitudinal breathing modes (see,
for example, Refs.~\cite{Thomas04} and~\cite{Grimm04} which deal with Fermi
gases close to unitarity).  These correspond to
$a_{s1} = a_{s2}$ and $a_{n1} = a_{n2}$.   
\section{Hydrodynamic Theory at T=0}
Since the variational principle given by
Eqs.~(\ref{L0c3}), (\ref{RR}), and 
(\ref{var2}) involves some fairly complex expressions, it
is useful to consider the case of zero temperature where the Lagrangian
simplifies
tremendously.  At $T=0$, the Landau two-fluid equations reduce to a single
hydrodynamic differential equation that was first used by Pitaevskii and
Stringari~\cite{StringariPit98} to
derive the corrections to the mean-field Gross-Pitaevskii
results~\cite{Stringari96} for
the collective mode frequencies in a dilute Bose gas.

At $T=0$, only the superfluid component persists and we have
$S=0$, $\rho_{n0} = 0$, and 
$\rho_{s0} = \rho_0$.  Eqs.~(\ref{L0c3})-(\ref{U}) then reduce to
\bea
\lefteqn{L^{(2)}[\bu_s] =}\nonumber\\&& \frac{1}{2}\int
d^3r\;\left\{\rho_{0}\bu_s^2\omega^2 -\left(\frac{\partial
\mu}{\partial\rho}\right)\left[\bnab\cdot\left(\rho_{0}\bu_s\right)\right]^2
\right\}.\nonumber\\\label{LT0} \eea 
From $\delta L^{(2)}/\delta\bu_s = \b0$, one obtains
\bea \omega^2\bu_s = -\bnab\left[\left(\frac{\partial
\mu}{\partial\rho}\right)\bnab\cdot\left(\rho_{0}\bu_s\right)\right]
\label{Euler0}. \eea Using the $T=0$ linearized continuity equation,
$\delta \rho =
-\bnab\cdot\left(\rho_{0}\bu_s\right)$, Eq.~(\ref{Euler0}) can be rewritten
as
\bea \omega^2\bu_s = \bnab\left[\left(\frac{\partial
\mu}{\partial\rho}\right)\delta\rho\right]. \label{euler}\eea 
Multiplying both sides of this expression by $\rho_0$ and taking the divergence,
we obtain a closed equation for the density fluctuations,
\bea \omega^2\delta\rho =
-\bnab\cdot\left[\rho_0\bnab\left\{\left(\frac{\partial
\mu}{\partial\rho}\right)\delta\rho\right\}\right], \label{euler2}\eea 
that is the basis of the $T=0$ 
hydrodynamic theory developed by Pitaevskii and Stringari~\cite{StringariPit98}.
Eq.~(\ref{euler2}) describes the low-energy collective modes of both
atomic Bose and two-component Fermi superfluid gases at $T=0$.  The only
difference between the
two quantum gases lies in the equation of state $\mu(\rho)$.  

It is evident that our variational approach will give expressions for the
collective mode frequencies at $T=0$ that agree with results derived from
solving 
Eq.~(\ref{euler2}) directly (see, for example, Ref.~\cite{StringariPit98}
for Bose gases, and Refs.~\cite{Stringari04EPL, Heiselberg04, Tosi04} for Fermi
gases close to unitarity). However, it is still useful to show explictly how
our formalism reproduces the $T=0$ results obtained in the recent literature.
As a specific application, we
consider the
breathing modes which were discussed at finite $T$ in   
the previous section.  The breathing modes of trapped Fermi gases
close to unitarity have been studied
extensively at $T=0$~\cite{Stringari04EPL, Heiselberg04, Zubarev04, Tosi04,
Bulgac05, Manini05,
Stringari05}, where
a simple ansatz for the density dependence of the chemical potential
$\mu(\rho)$ allows one to obtain simple expressions for the
frequencies of these modes.  We show
that in the limit of zero temperature, our general results for the breathing
mode frequencies at finite $T$, given by
Eqs.~(\ref{monopoles}) and (\ref{monopolen}), reproduce these well-known
expressions.

Since the normal fluid vanishes at zero temperature, we only have equations for
the superfluid component, given by
Eq.~(\ref{monopoles}), which reduce to \bea
M_{si}\omega^2 a_{si}=\sum_j k_{s,ij}a_{sj}\label{T=0eqn}\eea at $T=0$, 
where $k_{n}$ and $k_{sn}$ are zero.
The expression for $k_{s,ij}$, given by Eq.~(\ref{ksm}), becomes
\bea k_{s,ij} &=& \int d^{3}r\;\left(\frac{\partial
\mu}{\partial
\rho}\right)\frac{\partial(\rho_{0}x_i)}{\partial
x_i}\frac{\partial(\rho_{0}x_j)}{\partial x_j}, \label{ksmT=0} \eea
since $\rho_{s0} = \rho_0$ at $T=0$.

To obtain a simple analytic expression for the collective mode frequencies, the
following identity,
found by integration by parts, will prove useful:
\bea \lefteqn{\int d^{3}r\; \left(\frac{\partial \mu}{\partial
\rho}\right)\rho_0 x_j \frac{\partial \rho_0}{\partial x_j}=}\nonumber\\&& 
-\frac{1}{2}\int d^{3}r\; \rho_0^2 \Bigg\{ \left(\frac{\partial \mu}{\partial
\rho}\right) + 
x_j\frac{\partial}{\partial x_j}\left(\frac{\partial \mu}{\partial
\rho}\right)\Bigg\}. \label{useful4}\eea
Expanding out the derivatives in Eq.~(\ref{ksmT=0}) and using
Eq.~(\ref{useful4}), Eq.~(\ref{ksmT=0}) can be
rewritten as
\bea k_{s,ij} &=&\int
d^{3}r\;\Bigg\{\left(\frac{\partial \mu}{\partial \rho}\right)\frac{\partial
\rho_0}{\partial x_i}\frac{\partial \rho_0}{\partial x_j} x_i x_j \nonumber\\
&&-\frac{\rho_0^2}{2}\left(x_j\frac{\partial}{\partial x_j}
+x_i\frac{\partial}{\partial x_i}\right)\left(\frac{\partial
\mu}{\partial \rho}\right)\Bigg\}. 
\label{ksmT=02} \eea 
At $T=0$, Eq.~(\ref{useful3}) reduces to  
\bea\left(\frac{\partial \mu}{\partial \rho}\right) \frac{\partial
\rho_0}{\partial x_i} 
=-\omega_i^2x_i. \label{useful3T=0} \eea 
Using this result in Eq.~(\ref{ksmT=02}) and integrating by parts, we find \bea
\lefteqn{k_{s,ij} = \int d^{3}r\;\rho_0\Bigg\{(2\delta_{ij} + 1)\omega_i^2
x_i^2} \nonumber\\&&- 
\frac{\rho_0}{2}x_j\frac{\partial}{\partial x_j}\left(\frac{\partial
\mu}{\partial \rho}\right)
-\frac{\rho_0}{2}x_i\frac{\partial}{\partial x_i}\left(\frac{\partial
\mu}{\partial \rho}\right)\Bigg\}.
\label{ksmT=03} \eea

Assuming a so-called \textit{polytropic} equation of state (see, for example,
Refs.~\cite{Menotti02, Cozzini03, Tosi04, Heiselberg04,
Bulgac05,
Stringari05}),
\bea \mu[\rho] \propto \rho^{\gamma}, \label{eos}\eea
we find
\bea \rho_0\frac{\partial}{\partial x_i}\left(\frac{\partial \mu}{\partial
\rho}\right) &=& 
(\gamma - 1)\left(\frac{\partial \mu}{\partial \rho}\right)\frac{\partial
\rho_0}{\partial x_i}\nonumber\\
&=& -(\gamma - 1)\omega_i^2 x_i. \label{useful5}\eea
Substituting this relation into Eq.~(\ref{ksmT=03}), and making use of the
definition of the weighted mass $M_{si}$ given by Eq.~(\ref{breathingmasses}), 
we obtain the 
following expression for the ratio of the spring constant and the
weighted mass:
 \bea
\frac{k_{s,ij}}{M_{si}} =  2\omega_i^2 \delta_{ij} +
\left[\frac{(\gamma + 1)}{2}\omega_i^2 + \chi_{ji}
\frac{(\gamma - 1)}{2}\omega_j^2\right],\nonumber\\
\label{ksmT=04} \eea
where
\bea \chi_{ji} = \frac{\int d^{3}r\;\rho_0(\br) x_j^2}{\int
d^{3}r\;\rho_0(\br)x_i^2}.
\label{zeta} \eea

Using the Thomas-Fermi result $\mu_0 = \mu[\rho_0(\br)] +
U_{\text{ext}}(\br)$, the equilibrium density $\rho_0$ can be expressed as a
function of 
$U_{\text{ext}}(\br) = (1/2)(\omega_1^2x_1^2 + \omega_2^2 x_2^2 + \omega_3^2
x_3^2)$.  Making a change of variables, $x_i' = \omega_i x_i$, 
$\chi_{ji}$ becomes
\bea \chi_{ji} &=& \frac{\omega_i^2}{\omega_j^2}\frac{\int
d^{3}r'\;\rho_0[x_1'^2 + x_2'^2 + x_3'^2] x_j'^2}{\int d^{3}r'\;\rho_0[x_1'^2 +
x_2'^2 + x_3'^2] x_i'^2}\nonumber\\
&=& \frac{\omega_i^2}{\omega_j^2}. \label{zeta2} \eea
With this result, Eq.~(\ref{ksmT=04}) simplifies to
\bea
\frac{k_{s,ij}}{M_{si}}  = 2\omega_i^2 \delta_{ij} + \gamma\omega_i^2.
\label{ksmT=05} \eea  Combining this identity with 
Eq.~(\ref{T=0eqn}), we find \bea
\omega^2 a_{si}&=&\sum_j \left(2\omega_i^2 \delta_{ij}
+\gamma\omega_i^2\right)a_{sj}\nonumber\\
&=&2\omega_i^2a_{si}  + \gamma\omega_i^2\sum_ja_{sj}\label{T=0eqn5},\eea
which is equivalent to Eq.~(3) in
Ref.~\cite{Stringari05} when a polytropic equation of state is used. 

For an axisymmetric trap, $\omega_1 = \omega_2 = \omega_{\bot}$, $\omega_3 =
\omega_z$, and 
the axial and longitudinal breathing modes are characterized by solutions of the
form $a_1 = a_2 = a$.  In this
case, Eq.~(\ref{T=0eqn5}) gives the normal mode frequencies found by previous
authors~\cite{Stringari05, Cozzini03, Heiselberg04, Tosi04}, \bea
\omega^2  \!&=&\! \frac{1}{2}\Big[2(\gamma + 1)\omega_{\bot}^2 +
(\gamma + 2)\omega_z^2
\nonumber\\
&&\pm \sqrt{\left[2(\gamma + 1)\omega_{\bot}^2 - (\gamma +
2)\omega_z^2\right]^2 + 8\gamma^2\omega_z^2\omega_{\bot}^2}\Big].\nonumber\\
\label{Stringarieos} \eea

\section{Conclusions}
More than sixty years after Tisza~\cite{Tisza} and
Landau~\cite{Landau41}
proposed two-fluid hydrodynamics, their theory turns out to be very relevant in
the study of the dynamics of trapped Fermi superfluid gases with strong
interactions, close to unitarity. 
From linearized two-fluid hydrodynamics, one can obtain collective mode
frequencies in terms of equilibrium thermodynamic functions and
thermodynamic derivatives.  This is well known for uniform superfluids and
still occurs in the case of trapped gases, apart from the fact that the
equilibrium thermodynamic quantities are now dependent on position.  This means
that the two-fluid predictions are general, with a microscopic
theory only being needed for the elementary excitations
that determine the equilibrium thermodynamic functions.

Variational approaches have already been used extensively to obtain
hydrodynamic collective mode frequencies of superfluid Fermi gases
at $T=0$ close to
unitarity~\cite{Zubarev04, Bulgac05, Stringari05}, as well as boson-fermion
mixtures~\cite{Suzuki05}. In this paper, we have extended this approach
to
finite temperatures to deal with the coupled hydrodynamics of a superfluid and
normal fluid in a trapped gas. 

We have presented a variational method to determine the
collective mode frequencies of a trapped superfluid, based on an action of
the type first
proposed by Zilsel in 1950 to derive Landau two-fluid hydrodynamics.  By making
an
ansatz for the superfluid and normal fluid velocity fields, in this
formulation, the problem of
determing the collective mode frequencies is reduced to solving a set of
algebraic equations.  This
obviates the need to solve the
Landau two-fluid differential equations numerically, a difficult task in a
nonuniform trapped gas. 

We illustrated our formalism by considering a uniform gas, as
well as the
dipole and breathing modes for a harmonically confined gas.  To
make contact with recent discussions in the literature on trapped superfluid
Fermi gases, we also used our finite $T$
variational approach to discuss the collective
modes in the limiting case of a pure superfluid at
$T=0$~\cite{Stringari04EPL, StringariPit98}. 

In future work, we will apply the expressions derived in this paper and present
detailed numerical predictions for the frequencies of various hydrodynamic modes
in
trapped superfluid gases at finite temperatures.  In particular, we will
discuss
results in a Fermi superfluid for a strongly interacting molecular BEC at
finite $T$ (the BEC side of a Feshbach resonance).
 
In this paper, we have only discussed how to determine the normal modes of the
non-dissipative Landau two-fluid equations for a trapped (non-uniform)
superfluid.  The damping of these modes due to various hydrodynamic relaxation
processes can be calculated using the general expressions derived by Nikuni and
Griffin~\cite{NG04b}.

\begin{acknowledgments}
We would like to thank Duncan O'Dell for stimulating discussions about his
work on hydrodynamic modes in the unitarity region of Fermi superfluids.  We
also thank Eugene Zaremba for useful comments related to the
variational approach developed in Ref.~\cite{ZNGJLTP}.  Our research was
supported by a
grant from NSERC of Canada.   
\end{acknowledgments}

\appendix
\section{Comparison with Zilsel's Variational Principle}

As discussed in the text, our major interest in this paper is to find an
efficient way of
calculating the collective mode frequencies of a trapped superfluid in the
region where Landau's two-fluid hydrodynamics are valid.  Following
Zilsel~\cite{Zilsel50}, we did this in
Sections II-IV by
reformulating the two-fluid equations in terms of a variational theory
involving the action.  There is an extensive older literature
on
such a variational formulation in the context of superfluid helium.  In this
Appendix, we review in more detail some of the subtle issues which arose in
connection with Zilsel's seminal paper and their resolution.  

One criticism concerned Zilsel's use of  $x = \rho_n/\rho$ as an independent
variable which, it was argued, conflicted with the Landau
formalism~\cite{Lhuillier75}. Zilsel was ultimately
vindicated by Jackson who demonstrated that Zilsel's approach was completely
equivalent to those
proposed by some of his critics~\cite{Jackson78}. 
 
In our
approach to the two-fluid thermodynamics,
using the internal energy density given by Eq.~(\ref{dU}), we have deviated
slightly from the
variational method devised by Zilsel to derive the LTF equations. 
He considered the \textit{specific} internal
energy, which we denote by $U_Z$ to avoid confusion.  The two
energies are related by $U = \rho U_Z$.  Furthermore, while we have taken 
the internal energy density to be a function of the entropy $S$, the total
density ($\rho = \rho_n + \rho_s$), and the normal fluid
density $\rho_n$, Zilsel took
the specific internal energy to be a function of 
the entropy density $\sigma = S/\rho$, $\rho$, and the
ratio $x = \rho_n/\rho$.  However, one can show that   
the thermodynamics given in Section II, obtained from the Landau-Khalatnikov
energy
density, is identical to Zilsel's and our choice
of
independent variables is completely consistent with Zilsel's
choice.  
To compare our thermodynamic identities with Zilsel's, using Eqs.~(\ref{dU}) and
(\ref{Pressure}), with $U = \rho U_Z$ and hence, $dU = \rho dU_Z + U_Z d\rho$,
we find
\bea dU_Z = T d\sigma + \frac{P}{\rho^2}d\rho + \frac{1}{2}(\bv_n -
\bv_s)^2 dx ,
\label{dUZ} \eea  where $\sigma = S/\rho$ and $x = \rho_n/\rho$.  This
relation gives precisely the thermodynamic relations used by Zilsel in his
derivation of the LTF equations.  As well, Eq.~(\ref{dUZ}) was derived by
Jackson using thermodynamic arguments~\cite{Jackson78}.  Zilsel
chose to use $\sigma, \rho$, and $x$ as independent variables as
these differentials appear in Eq.~(\ref{dUZ}).  In contrast, with
the internal energy density given by Eq.~(\ref{dU}), the differentials
of $S$,
$\rho$, and $\rho_n$ are the natural choice for independent variables.  

A more valid criticism directed at Zilsel's work concerned the fact that, while
he obtained the full LTF equations, his
variational principle forced the normal fluid velocity to be irrotational when
$S/\rho_n$  is a constant.  Thus, Zilsel's variational principle only describes
a subset of all possible solutions of the LTF equations.  
Following an argument originally due to Lin, this deficiency can be corrected
by introducing an additional constraint, given by Eq.~(\ref{circcont}), into the
variational principle that
accounts for conservation of circulation (or vorticity) in the normal
fluid~\cite{
Lhuillier75, Geurst76, Jackson78, Geurst80,Purcell81, Yourgau}.  To see this
explicitly, consider Eq.~(\ref{nabphi}).
Without including the
circulation constraint, equivalent to setting 
$\eta=0$, we recover Zilsel's expression for the normal
fluid velocity.  It is apparent from Eq.~(\ref{nabphi}) that for the case when
the
entropy divided by the normal fluid density is uniform, the normal
fluid velocity is
irrotational.  By including the circulation constraint, the difference between
the normal and superfluid velocities multiplied by $\rho_n/S$ assumes the
form of a Clebsch transformation~\cite{Lamb}, with the local functions $\eta$
and $\gamma$
acting as the
Clebsch
potentials.  In this case, the normal fluid velocity is no longer
restricted.

\section{ZNG vs. ZGN hydrodynamics}
We briefly summarize some key features of the derivation of two-fluid
hydrodynamics for a trapped Bose gas.  As discussed in detail in
Refs.~\cite{ZNGJLTP}
and~\cite{NG01}, the 
ZNG microscopic coupled equations involve in a crucial way a term $\Gamma_{12}$
 which is related to the collisionless transfer of atoms between the condensate
and thermal cloud.  In the linearized ZNG formalism, one can show that in local
equilibrium, 
\bea \delta\Gamma_{12} = -\frac{n_{c0}}{k_B
T\tau_{12}}\delta\mu_{\text{diff}}. \label{deltagamma12} \eea
This involves the difference in the chemical potentials of the condensate
($\mu_c$) and the thermal cloud ($\tilde{\mu}$)
\bea \mu_{\text{diff}} = \tilde{\mu} - \mu_c. \label{mudiff} \eea
In static equilibrium, $\Gamma_{12}$ vanishes and $\tilde{\mu}_0 = \mu_{c0}$,
assuming that $\bv_{c0} = \bv_{n0} = \b0$.  In Eq.~(\ref{deltagamma12}), the
relaxation 
time $\tau_{12}$ describes collisions between atoms in the condensate and
thermal gas components.   The relaxation time $\tau_{\mu}$ introduced in
Ref.~\cite{ZNGJLTP} which determines the rate at which local diffusive
equilibrium is achieved ($\tilde{\mu} = \mu_c$), is given by $\tau_{\mu} =
\sigma_H \tau_{12}$, where $\sigma_H$ is a dimensionless hydrodynamic
normalization factor defined in Ref.~\cite{ZNGJLTP} involving local
equilibrium thermodynamic functions.  

ZNG derive the non-dissipative Landau two-fluid equations in the limit of
$\omega\tau_{\mu} \ll 1$.  In this limit, one finds $\delta\mu_{\text{diff}}
\rightarrow 0$, but $\delta\Gamma_{12}$ in Eq.~(\ref{deltagamma12})
remains finite and is given by \bea \delta\Gamma_{12}(\br,t) \!=\!
\sigma_H\!\left\{\bnab\cdot\left[n_c(\bv_c - \bv_n)\right] +
\frac{1}{3}n_c\bnab\cdot\bv_n\right\}. \label{deltagammadef}\eea
As discussed in Ref.~\cite{ZNGJLTP}, this source term in the continuity
equations Eqs.~(\ref{rhoncont}) and (\ref{rhoscont}) plays a crucial role in
establishing the equivalence between
the ZNG equations (in the limit that $\omega\tau_{\mu}\rightarrow 0$) and the
standard Landau two-fluid equations considered in the text of this paper. 
Thus, $\Gamma_{12}$ is a concrete example of the source term $\Gamma$
($=m\Gamma_{12}$) which arises in Eqs.~(\ref{vnt}) and (\ref{rhoncont}), in the
context of a dilute Bose gas.  

This is a convenient place to note that in Refs.~\cite{ZNGPRA98, ZG97, ZNGJLTP},
another version of the two-fluid hydrodynamics was discussed
in which the collisional transfer of atoms between the condensate and thermal
cloud was ignored.  This corresponds to setting $\Gamma_{12} = 0$ or,
equivalently, the limit $\omega\tau_{\mu} \gg 1$.  This ``ZGN" two-fluid
hydrodynamics was used to calculate the temperature-dependence of the
frequencies of first and second sound in a weakly-interacting uniform Bose
gas~\cite{ZG97}.
In this case, the velocities were almost identical (see Fig.1 of
Ref.~\cite{ZNGJLTP}) to those based on the Landau two-fluid
hydrodynamics ($\omega\tau_{\mu} \ll 1$ limit).  Expressions for the two sound
velocities based on the LTF equations are given in Appendix C of
Ref.~\cite{NG01},
showing the small corrections which depend on the factor $\sigma_H$ in
Eq.~(\ref{deltagammadef}). 

The variational theory developed in Ref.~\cite{ZNGJLTP} for determining the
normal mode oscillations in a trapped gas was worked out originally for
the ZGN two-fluid hydrodynamics~\cite{ZNGPRA98}.  The results for monopole,
dipole, and
quadrupole oscillations are summarized in Fig.~1 of Ref.~\cite{ZNGPRA98} and
Figs. 4 and 5 of Ref.~\cite{ZNGJLTP}.  For the dipole modes, the frequencies are
identical to Eqs.~(\ref{firstsound}) and (\ref{secondsound}), but with 
\bea k_{sn} = -2g\int d^{3}r\;\frac{\partial n_{c0}}{\partial
x_i}\frac{\partial \tilde{n}_0}{\partial x_i}, \label{ZNGdipole}\eea
where $g = 4\pi\hbar^2a/m$.  It can be shown, however, making use of
approximations consistent with
those used to derive ZGN hydrodynamics, that Eq.~(\ref{ksndipole}) reduces to
Eq.~(\ref{ZNGdipole}).  The details of this derivation will be given elsewhere.
 
The variational approach of Ref.~\cite{ZNGJLTP} was formally extended to deal
with the ZNG
hydrodynamics in Ref.~\cite{NG04b}, but it was not used to calculate the
collective mode frequencies.  Instead, the out-of-phase mode was approximated
by extending the $T=0$ pure condensate mode to finite $T$ by using a reduced
condensate density.  The in-phase solution was approximated at finite $T$ by
the thermal cloud oscillation above $T_c$, but with a reduced thermal cloud
density.  As discussed in Section VII of Ref.~\cite{ZNGJLTP}, these are often
good first
approximations.  


\begin{thebibliography}{99}
\bibitem{note1} We have not considered the possibility that $U$
also depends on the gradient of the superfluid density $\bnab\rho_{s}$.  Its
inclusion would lead to additional terms in the variational
principle and hence, to a modified form of the LTF equations. This
approximation is
equivalent to ignoring the ``quantum pressure" contribution to the energy. 
Geurst~\cite{Geurst80} included such contributions in his extension of
Zilsel's method.
\end{thebibliography}

\end{document}